\documentstyle[aps,prl,preprint,tighten]{revtex}
\newcommand{\half}{{\textstyle{1\over2}}}
\newcommand{\halfhalf}{(\half,\half)}
\newcommand{\D}{{\mathcal D}}
\newcommand{\tfrac}[2]{{\textstyle{\frac{#1}{#2}}}}
\newcommand{\Kappa}{\kappa}
\renewcommand{\d}{{\rm d}}
\newcommand{\Tr}{\mbox{Tr\,}}
\newcommand{\Det}{\mbox{{\rm Det}}}
\newcommand{\beq}{\begin{equation}}
\newcommand{\eeq}[1]{\label{#1}\end{equation}}
\newcommand{\bea}{\begin{eqnarray}}
\newcommand{\eea}[1]{\label{#1}\end{eqnarray}}

\newcommand{\rtitle}[1]{\ifpreprintsty{\sl #1},\fi}

\begin{document}
\preprint{\vtop{\hbox{MCTP-01-06}\hbox{hep-th/0102093}\vskip24pt}}

\title{Quantum discontinuity between zero and infinitesimal graviton
mass with a $\Lambda$ term}

\author{F.~A.~Dilkes, M.~J.~Duff, James T.~Liu and
H.~Sati\footnote{Email addresses:
\{fadilkes, mduff, jimliu, hsati\}@umich.edu}}

\address{Michigan Center for Theoretical Physics\\
Randall Laboratory, Department of Physics, University of Michigan\\
Ann Arbor, MI 48109--1120, USA}

\maketitle

\begin{abstract}
We show that the recently demonstrated absence of the usual discontinuity for 
massive spin 2 with a $\Lambda$ term is an artifact of the tree approximation, 
and that the discontinuity reappears at one loop.
\end{abstract}


\ifpreprintsty\else
\begin{multicols}{2}
\fi
\narrowtext

%%%%%%%%%%%%%%%%%%%%%%%%%%%%%%%%%%%%%%%%%%%%%%%%%%%%%%%%%%%%%%%%%%%%%%%%%%%%%%
%% Introduction

An old question is whether the graviton has exactly zero mass or
perhaps a small but non-zero mass.  This issue seemed to have been
resolved by van Dam and Veltman \cite{vdv} and, independently,
Zakharov \cite{zak} when they noted that there is a discrete difference
between the propagator of a strictly massless graviton and that of a
graviton with mass $M$ in the $M\rightarrow 0$ limit.  This difference
gives rise to a discontinuity between the corresponding amplitudes
involving graviton exchange.  In particular, the bending of light by the
sun in the massive case is only 3/4 of the experimentally confirmed massless
case, thus ruling out a massive graviton.

Recently, however, the masslessness of the graviton has been called
into question by two papers \cite{Kogan,Porrati} pointing out that the
van Dam-Veltman-Zakharov discontinuity disappears if, instead of being
Minkowski, the background spacetime is anti-de Sitter (AdS).  The same
result in de Sitter space had earlier been obtained in \cite{Higds,Hig}.
In fact, as shown below, this can be extended to any Einstein space
satisfying
\begin{equation}
\label{Einstein}
R_{\mu\nu} = \Lambda g_{\mu\nu}
\end{equation}
with a non-zero cosmological constant $\Lambda\neq0$ provided
$M^2/\Lambda\to0$.

These results remain surprising, however, since the massive graviton
retains five degrees of freedom, while the massless one only has two.
Although these extra states decouple from a covariantly conserved
stress tensor for $M^2/\Lambda\to0$, yielding a smooth classical
limit, they are nevertheless still present in the theory, suggesting
that a discontinuity may remain at the quantum level.  In
this letter, we demonstrate that this is indeed the case by
calculating the one loop graviton vacuum amplitude for a massive
graviton and showing that it does not reproduce the result for the
massless case in the limit.  Thus the apparent absence of the
discontinuity is only an artifact of the tree approximation and the
discontinuity reappears at one loop.

%%%%%%%%%%%%%%%%%%%%%%%%%%%%%%%%%%%%%%%%%%%%%%%%%%%%%%%%%%%%%%%%%%%%%%%%%%%%%%
%% Massive gravity (Stuckelberg formalism)

In order to handle a massive graviton and resulting loss of general
covariance, we employ the alternative ``St\"uckelberg''
formalism~\cite{stuck,Hig}.  This formalism reproduces a tree-level
amplitude for
conserved sources which agrees with Ref.~\cite{Porrati} for general
$\Lambda$ and $M$, but confirms the naively expected discontinuity
in the determinant describing the one-loop effective action for the
background configuration.

We work in four dimensions with Euclidean signature
\mbox{$({}+{}+{}+{}+{})$}.
As in Ref.~\cite{Porrati}, our starting point is the action
\begin{equation}
S[h_{\mu\nu},T_{\mu\nu}] = S_L[h_{\mu\nu}] + S_M[h_{\mu\nu}]
+ S_T[h\cdot T]\, ,
\end{equation}
where $S_L$ is the Einstein-Hilbert action with cosmological constant
$S_E = -\frac{1}{16\pi G}
\int\d^4 x \sqrt{\hat g} (\hat R - 2 \Lambda)$, linearized about a
background metric $g_{\mu\nu}$ satisfying Eq.~(\ref{Einstein}) according
to $\hat g_{\mu\nu} = g_{\mu\nu} + \Kappa h_{\mu\nu}$ where
$\kappa^2 = 32 \pi G$.  For such a background, this linearized action for
$h_{\mu\nu}$ is
\begin{eqnarray}
\label{SL}
S_L = \int \d^4 x \sqrt{g}
\Bigl[&&\tfrac{1}{2}\tilde h^{\mu\nu}
\left(-g_{\mu\rho}g_{\nu\sigma}\Box-2R_{\mu\rho\nu\sigma}\right)h^{\rho\sigma}
\nonumber\\
&& -\nabla^\rho \tilde h_{\rho\mu} \nabla^{\sigma} \tilde h_\sigma{}^\mu
\Bigr] \, ,
\end{eqnarray}
where $\tilde h_{\mu\nu} = h_{\mu\nu} - \frac{1}{2} g_{\mu\nu}
h^{\sigma}{}_\sigma$.  The linearized Lagrangian $S_L$ has a gauge
symmetry described by a vector $\xi_\mu(x)$
\begin{equation}
\label{lineargauge}
h_{\mu\nu} \rightarrow h_{\mu\nu}  + 2
\nabla_{(\mu} \xi_{\nu )}\, .
\end{equation}
All indices are raised and lowered with respect to the metric $g_{\mu\nu}$
and $\nabla_\mu$ is taken to be covariant with $\nabla_\mu g_{\lambda\sigma}
= 0$.  The Pauli-Fierz spin-2 mass term
\begin{equation}
S_M = \tfrac{M^2}{2} \int \d^4x \sqrt{g}
\left[ h^{\mu\nu} h_{\mu\nu} - (h^\mu{}_\mu)^2 \right]
\end{equation}
breaks the symmetry (\ref{lineargauge}).  Finally, the source term is
given by
\begin{equation}
S_T = \int d^4 x \sqrt{g} \, h_{\mu\nu} T^{\mu\nu} \, .
\end{equation}
As in Ref.~\cite{Porrati}, we apply the simplifying assumption that
$T_{\mu\nu}$ is conserved with respect to the background metric,
$\nabla_{\mu} T^{\mu\nu} = 0$, although it will become clear how
this assumption may be relaxed, if needed.

In addition to reproducing the classical correlators for the sources
$T_{\mu\nu}$ given in Ref.~\cite{Porrati}, we would also like to determine
whether the one-loop effective action for the background field configuration
$g_{\mu\nu}$ is continuous in the $M^2/\Lambda \rightarrow 0$
limit.  Since many of the relevant calculations have already been given for
the propagation of massless $M^2=0$ gravitons in Ref.~\cite{Christensen},
we now proceed directly to the case $M^2 \neq 0$.

We employ the path-integral (PI) formalism and define
the generating functional
\begin{equation}
\label{Z}
Z[g,T] = \int \D h \, e^{-\left( S_L[h] + S_M[h] + S_T[h\cdot T]\right)}\, .
\end{equation}
Since this theory no longer contains a gauge symmetry of the form
(\ref{lineargauge}), one could, in principle, proceed directly by
calculating the propagators and determinants associated with the
quadratic form implied by $S_L + S_M$ without the appearance of any
ghost sector.  Indeed the propagators were computed directly in
Ref.~\cite{Porrati} using a suitable component decomposition.  Instead,
we introduce the gauge symmetry (\ref{lineargauge}) using a
St\"uckelberg \cite{stuck} formulation. This will help to make contact
with the operators appearing in Ref.~\cite{Christensen} for the pure
massless case.

%% Step 1

In the St\"uckelberg formalism, the gauge symmetry of the massless
theory is restored by introducing an auxiliary vector field $V_\mu$.
We first multiply $Z[g,T]$ by an integration $\int \D V$ over all
configurations of this decoupled field, and then perform the shift
$h_{\mu\nu} \rightarrow h_{\mu\nu} - 2M^{-1} \nabla_{(\mu}  V_{\nu )}$.
Since $S_L$ and $S_T$ are gauge invariant in themselves, the only effect
of this shift is to make the replacement
\begin{equation}
S_M[h_{\mu\nu}]\to S_M[h_{\mu\nu} - 2 M^{-1} \nabla_{(\mu} V_{\nu )}]
\end{equation}
in (\ref{Z}).  Thus $S_M$ becomes a ``St\"uckelberg mass'', and gauge
invariance is restored under the simultaneous shift
\begin{equation}
\label{lineargauge2}
V_\mu \rightarrow V_\mu + M \xi_\mu \, ,
\end{equation}
along with the original transformation (\ref{lineargauge}).

%% Step 2

We now gauge fix by identifying $V$ with the longitudinal part of
$\tilde h$, namely $M V_\mu = \nabla^\rho \tilde h_{\rho \mu}$.
This choice is made in order to simplify the relevant operators
appearing in the action, and is accomplished by adding to the action
the gauge-fixing term
\begin{equation}
S_{\rm gf} = \int \d^4 x \sqrt{g}
\left( \nabla^\rho \tilde h_{\rho\mu} - M V_{\mu} \right)
\left( \nabla^\rho \tilde h_\rho^\mu - M V^{\mu} \right)\, .
\end{equation}
This has the effect of canceling the last term in equation (\ref{SL})
in addition to the cross-term in $S_M$ between $V$ and (the traceless
component of) $h_{\mu\nu}$.

To properly account for interactions with the background metric, we must
also include a Faddeev-Popov determinant connected with the variation of
this gauge condition under (\ref{lineargauge}) and (\ref{lineargauge2}).
It is straightforward to show that the appropriate determinant is
\begin{equation}
\label{vectdet}
\Det \left[(-\Box + M^2) \delta_\mu^\lambda - R_\mu^\lambda \right]
=\Det \left[ \Delta\halfhalf - 2 \Lambda + M^2 \right]\, ,
\end{equation}
where the second-order vector spin operator is defined by
$\Delta\halfhalf \xi_\mu \equiv - \Box \xi_\mu + R_{\mu\nu} \xi^\nu$
\cite{Christensen} and we have exploited the Einstein condition
(\ref{Einstein}) for the background metric.

%% Step 3

After gauge fixing, there remains a coupling proportional to
$h^\sigma{}_\sigma \nabla \cdot V$
which can be eliminated by making
the change of variables $V_\mu \rightarrow V_\mu + \alpha
\nabla_\mu h^\sigma{}_\sigma$.  It turns out that the remaining
component $h_{\mu\nu}$ becomes completely decoupled from $V_\mu$
if we choose $\alpha = M/(-4 \Lambda + 2M^2 )$.
This shift is a convenient but
not
crucial step in our calculation; our results will also hold
in the apparently singular case when
$2 \Lambda= M^2 \neq 0$.

To highlight the tensor structure of the gauge-fixed action, we
decompose the metric fluctuation $h_{\mu\nu}$ into its traceless and
scalar parts: $\phi_{\mu\nu} \equiv h_{\mu\nu} - \frac{1}{4} g_{\mu\nu}
h^\sigma{}_\sigma$, and $\phi \equiv h^\sigma{}_\sigma$.  The source may
similarly be split into its irreducible components $j_{\mu\nu}$ and $j$,
so that $T_{\mu\nu} = j_{\mu\nu} + \frac{1}{4} g_{\mu\nu} j $.
The gauge-fixed action then becomes
\begin{eqnarray}
\label{Sint}
\tilde S
& = & \int \d^4 x \sqrt{g} \Bigl[
\tfrac{1}{2} \phi^{\mu\nu} \left( \Delta(1,1)
- 2 \Lambda + M^2 \right) \phi_{\mu\nu} \nonumber \\
& & \hspace{1.6cm}
- \tfrac{1}{8} \left( \tfrac{-2 \Lambda + 3 M^2}{-2 \Lambda + M^2} \right)
 \phi \left( \Delta(0,0) -
2 \Lambda + M^2 \right) \phi \nonumber\\
& & \hspace{1.6cm}
+ V^\mu (\Delta\halfhalf -2 \Lambda + M^2) V_\mu
- (\nabla \cdot V)^2
\nonumber \\
& & \hspace{1.6cm}
+ \phi_{\mu\nu} j^{\mu\nu} + \tfrac{1}{4} \phi j \Bigr] \, .
\end{eqnarray}
The second-order spin operators are the scalar Laplacian
$\Delta(0,0) \equiv - \Box$ and the Lichnerowicz operator
for symmetric rank-2 tensors
$\Delta(1,1) \phi_{\mu\nu} = - \Box \phi + R_{\mu\tau}\phi^\tau_\nu
+ R_{\nu\tau} \phi_\mu^\tau - 2 R_{\mu\rho\nu\tau}\phi^{\rho\tau}$
\cite{Christensen}.

At this point, the tree-level amplitude for the current $T_{\mu\nu}$
can be read from the action (\ref{Sint}) directly and is given by
\begin{eqnarray}
A[T] & = & \tfrac{1}{4} \Bigl[
2\, j^{\mu\nu} \left(\Delta(1,1)-2\Lambda+M^2 \right)^{-1} j_{\mu\nu}\\
& & \hspace{0.3cm}
-\tfrac{1}{2} \left(\tfrac{-2 \Lambda + M^2}{-2 \Lambda + 3M^2}\right)
j \left(\Delta(0,0)-2\Lambda+M^2\right)^{-1} j  \Bigr] \, ,\nonumber
\end{eqnarray}
since there are no sources for $V_{\mu}$.
Writing $j_{\mu\nu}$ and $j$ in terms of $T_{\mu\nu}$, we find that
\begin{eqnarray}
\label{amplitude}
A[T] & = & \tfrac{1}{4} \Bigl[
2 \, T^{\mu\nu} \left(\Delta(1,1)-2\Lambda + M^2 \right)^{-1} T_{\mu\nu}\\
& & \hspace{0.3cm}
-\left(\tfrac{-2 \Lambda + 2 M^2}{-2 \Lambda + 3 M^2 } \right)
T^\mu_\mu \left(\Delta(0,0) -2 \Lambda + M^2 \right)^{-1} T^\mu_\mu
\Bigr] \, , \nonumber
\end{eqnarray}
in agreement with the result of Ref.~\cite{Porrati},
modulo an overall convention-dependent factor of $1/4$;
(the removable singularity at $\Box= -4 \Lambda/3$
reported in Ref.~\cite{Porrati} is absent in this formulation).
We note here that there would be sources for the St\"uckelberg fields
if one were to relax the assumption of a conserved stress tensor.
In this case, one needs only to account for the shifts in $h_{\mu\nu}$
and $V_\mu$ to see how $T_{\mu\nu}$ contributes to sources
for $V_\mu$ (and $\chi$ below).

%% Step 4

Together, (\ref{Sint}) and (\ref{vectdet}) provide a representation of $Z$
with no manifest vector gauge symmetry.  The terms involving $V_\mu$ in
(\ref{Sint}) correspond to a massive Maxwell action in the Einstein
background with an effective mass $m^2 = M^2-2\Lambda$ which breaks the
would-be $U(1)$ invariance whenever $m^2 \ne 0$.  As before, we impose
gauge symmetry in the St\"uckelberg formalism by introducing a
scalar field $\chi$ and making the change of variables
$V_\mu \rightarrow V_\mu - M^{-1} \nabla_\mu \chi $.  Only the penultimate
line of (\ref{Sint}) changes as a result of this shift.

%% Step 5

By construction, the resulting action is now invariant  under the
gauge transformation
\begin{eqnarray}
V_\mu & \rightarrow & V_\mu + \nabla_\mu \zeta\, , \nonumber\\
\chi & \rightarrow & \chi + M \zeta \, .
\end{eqnarray}
One can then choose a gauge-condition to simplify the shifted action.
It is useful to associate the longitudinal component of $V$ with $\chi$
according to $M\nabla \cdot V = (-2 \Lambda + M^2) \chi$.  This is done
by the addition of a gauge-fixing term
\begin{equation}
\label{eq:gf2}
S_{\rm gf}' = \int \d^4 x \sqrt{g}
\left(\nabla \cdot V - \tfrac{-2 \Lambda + M^2}{M} \chi \right)^2 \, ,
\end{equation}
along with a corresponding scalar Faddeev-Popov determinant
\begin{equation}
\label{FPprimed}
\Det\left[\Delta(0,0) - 2 \Lambda + M^2\right] \, .
\end{equation}
The $(\nabla \cdot V)^2$ and $\chi \nabla \cdot V$ terms in
(\ref{eq:gf2}) are designed to cancel against corresponding terms in the
shifted version of (\ref{Sint}).  If any $\phi \nabla \cdot V$
coupling had remained in Eq.~(\ref{Sint}), (for instance, if we had not
performed the shift $V_\mu \rightarrow V_\mu + \alpha \nabla_\mu \phi$),
then this too could have been eliminated by incorporating
an appropriate $\phi$ dependence in the gauge-fixing (\ref{eq:gf2}).
The additional mixing between
$\phi$ and $\chi$ would present no difficulty.
Hence, the results (if not the method) presented here
also hold if $M^2=2\Lambda$.  The case $M^2=2\Lambda/3$ must be
treated separately \cite{DLS}, however, since from (\ref{Sint}) the trace mode
completely decouples, leaving only four degrees of freedom instead of
five \cite{deser4}. 

The final completely gauged-fixed action is now given by (\ref{Sint})
with $-(\nabla\cdot V)^2$ replaced by the quadratic scalar term
\begin{equation}
\tfrac{-2 \Lambda + M^2 }{M^2}
\chi \left(\Delta(0,0) - 2 \Lambda + M^2 \right) \chi \, .
\end{equation}
Along with the addition to the two Faddeev-Popov determinants (\ref{vectdet})
and (\ref{FPprimed}), this provides a complete description of $Z$, including
couplings to the background metric.
We can integrate over all species to find the first quantum correction
\ifpreprintsty\else
\end{multicols}
\narrowtext
\hbox to \hsize{\hrulefill}
\fi
\widetext
\begin{eqnarray}
Z[g,T]  \propto
e^{-A[T]}
&&\Det \Bigl[\Delta\halfhalf - 2 \Lambda + M^2 \Bigl]
\Det \Bigl[\Delta(0,0) - 2 \Lambda + M^2 \Bigl]
\Det \Bigl[\Delta(1,1) - 2 \Lambda + M^2 \Bigl]^{-1/2}
%% equation formatting %%
\ifpreprintsty\kern-10pt\fi
\nonumber\\
\times&&\Det \Bigl[\Delta\halfhalf - 2 \Lambda + M^2 \Bigl]^{-1/2}
\Det \Bigl[\Delta(0,0) - 2 \Lambda + M^2 \Bigl]^{-1/2}
\Det \Bigl[\Delta(0,0) - 2 \Lambda + M^2 \Bigl]^{-1/2} \kern-1.5em ,
%% equation formatting %%
\ifpreprintsty\kern-37pt\nonumber\\\fi
\end{eqnarray}
where the operator $\Delta(1,1) -2 \Lambda + M^2$ arises in the
traceless sector $\phi^{\mu\nu}$ sector so its determinant refers
to traceless modes only.

In addition to the propagator~(\ref{amplitude}),
this allows us to compute the
one-loop contribution
\begin{equation}
\Gamma^{(1)}[g] = - \ln Z[g,0]
= -\tfrac{1}{2} \ln \Det \Bigl[\Delta\halfhalf - 2 \Lambda + M^2 \Bigl]
+ \tfrac{1}{2} \ln \Det \Bigl[\Delta(1,1) - 2 \Lambda + M^2 \Bigl]
\end{equation}
to the effective action for the Einstein background $g_{\mu\nu}$. This is
now to be compared with the one loop contribution in the strictly massless
case \cite{Christensen}
\begin{equation}
\Gamma^{(1)}[g] = - \ln Z[g,0]
= - \ln \Det \Bigl[\Delta\halfhalf - 2 \Lambda  \Bigl]
+ \tfrac{1}{2} \ln \Det \Bigl[\Delta(1,1) - 2 \Lambda  \Bigl]
+ \tfrac{1}{2} \ln \Det \Bigl[\Delta(0,0) - 2 \Lambda \Bigl]
%% equation formatting %%
\ifpreprintsty\kern-14pt\fi
\end{equation}
\ifpreprintsty\else
\narrowtext \dimen0\hsize
\widetext
\hbox to \hsize{\hss \hbox to \dimen0{\hrulefill}}
\begin{multicols}{2}
\fi
\narrowtext
The difference in these two expressions reflects the fact that 5
degrees of freedom are being propagated around the loop in the massive
case and only 2 in the massless case. Denoting the dimension of the
spin $(A,B)$ representation by $D(A,B)=(2A+1)(2B+1)$,
we count $D(1,1)-D(1/2,1/2)=5$ for the massive case, while
$D(1,1)-2D(1/2,1/2)+D(0,0)=2$ for the massless one.

It remains to check that there is no conspiracy among the eigenvalues of
these operators that would make these two expressions coincide. To show
this, it suffices to calculate the coefficients in the heat-kernel
expansion for the graviton propagator associated with $S_L + S_M$,
and compare it with the massless case given in Ref.~\cite{Christensen}.
The coefficient functions $b_k^{(\Lambda)}$ in the expansion
\begin{equation}
\Tr e^{-\Delta^{(\Lambda)} t}
= \sum_{k=0}^\infty t^{(k-4)/2}
\int \d^4x \sqrt{g} \, b_k^{(\Lambda)}
\end{equation}
were calculated in Ref.~\cite{Christensen} for general ``spin operators''
$\Delta^{(\Lambda)}(A,B)\equiv\Delta(A,B) -2\Lambda$ with the result
\begin{eqnarray}
180(4\pi)^2 b_4^{(\Lambda)}(1,1)&=&189R_{\mu\nu\rho\sigma}
R^{\mu\nu\rho\sigma}-756\Lambda^{2}\, ,功nonumber\\
180(4\pi)^2 b_4^{(\Lambda)}(\half,\half)&=&-11R_{\mu\nu\rho\sigma}
R^{\mu\nu\rho\sigma}+984\Lambda^{2}\, ,功nonumber\\
180(4\pi)^2 b_4^{(\Lambda)}(0,0)&=&R_{\mu\nu\rho\sigma}
R^{\mu\nu\rho\sigma}+636\Lambda^{2}功, .
\end{eqnarray}
It is straightforward to extend those results to relevant massive
operators $\Delta^{(\Lambda,M)}(A,B) \equiv \Delta(A,B) - 2 \Lambda + M^2$.
The coefficients $b_k^{(\Lambda,M)}(A,B)$
for these operators are perfectly smooth functions
of $M^2$. Thus, as $M^2 \rightarrow 0$, we obtain
\begin{eqnarray}
180(4\pi)^2b_4^{(\Lambda,M)}&&({\rm total})\nonumber\\
&&= 180(4\pi)^2 \left[
b_4^{(\Lambda,M)} (1,1) - b_4^{(\Lambda,M)}\halfhalf \right]\nonumber\\
&& \rightarrow
 200 R_{\mu\nu\rho\sigma} R^{\mu\nu\rho\sigma} -1740 \Lambda^2 \, ,
\end{eqnarray}
which clearly differs from the $M^2=0$ result
\begin{eqnarray}
180(&&4\pi)^2b_4^{(\Lambda)}({\rm total})\nonumber\\
&&= 180 (4\pi)^2 \left[b_4^{(\Lambda)}(1,1)-2b_4^{(\Lambda)}(1/2,1/2)
+b_4^{(\Lambda)}(0,0)\right]\nonumber\\
&&= 212 R_{\mu\nu\rho\sigma} R^{\mu\nu\rho\sigma} - 2088 \Lambda^2 \, .
\end{eqnarray}
(These one-loop differences between massive and massless spin 2 in
the $\Lambda=0$ case are well-known \cite{CD}).
Even in the case of backgrounds with constant curvature
\begin{eqnarray}
R_{\mu\nu\rho\sigma}&=&
\tfrac{1}{3}\Lambda(g_{\mu\nu}g_{\rho\sigma}-g_{\mu\rho}g_{\nu\sigma})
\, ,\nonumber\\
R_{\mu\nu\rho\sigma}R^{\mu\nu\rho\sigma}&=&\tfrac{8}{3}\Lambda^{2}功, ,
\end{eqnarray}
there is no cancellation. Thus we conclude that the absence of the
discontinuity between the $M^2 \rightarrow 0$ and $M^2=0$ results for
massive spin 2, demonstrated in Ref.~\cite{Porrati,Kogan}, is an artifact of
the tree approximation and that the discontinuity itself persists at one
loop.

%% Conclusion

That the full quantum theory is discontinuous is not surprising
considering the different degrees of freedom for the two cases.
However, as seen in (\ref{Sint}), the three additional longitudinal
degrees of freedom of the massive graviton do not couple to a conserved
stress tensor.  Thus, in the absence of any self-couplings (or at
tree-level), the additional longitudinal modes would decouple from
matter, yielding a smooth $M^2\to 0$ limit.  Nevertheless, due to
these self-couplings (seen here as couplings to the background metric
in the linearized approach), these additional modes do not decouple,
thus yielding the resulting discontinuity in the massless limit.
(This result also suggests that the situation would be similar for the
spin-$\frac{3}{2}$ case considered in Ref.~\cite{vanN,Deser}.)  Of 
course, these one loop effects are very small and so experiments such as the 
bending of light would still not be able to 
distinguish a massless graviton from a very light graviton in the 
presence of a non-zero cosmologocal constant.

We finish with the important caveat that the $M\rightarrow0$ discontinuity
for fixed $\Lambda$ of the massless limit of massive spin-2 we have
demonstrated applies to fields described by the action appearing in
(\ref{Z}) discussed in Ref.~\cite{Porrati,Kogan}.  One may question 
whether this is a suitable action to describe the interaction of massive 
gravitons.  We are not necessarily
ruling out a smooth limit for other actions that might appear in Kaluza-Klein
or brane-world models, for example.  Indeed one would
expect a smooth limit if the mass is acquired spontaneously
\cite{Duff} rather than through an explicit Pauli-Fierz term. In
conventional Kaluza-Klein models, however, this limit, though smooth, would
also be the decompactification limit and would result in massless
gravitons in the higher dimension rather than four dimensions.
A closer examination would be necessary to discern the form of the
effective action describing the trapped graviton of the brane-world
scenario of Refs.~\cite{bigrav,karch}.

\section*{Acknowledgements}
FAD would like to extend thanks to D.G.C.~McKeon for fruitful suggestions
and discussions, and to the Natural Sciences and Engineering Research Council
of Canada (NSERC) for support.  We wish to thank R.~Akhoury and S.~Deser for
enlightening discussions.  This research was supported in part by DOE Grant
DE-FG02-95ER40899.

%%%%%%%%%%%%%%%%%%%%%%%%%%%%%%%%%%%%%%%%%%%%%%%%%%%%%%%%%%%%%%%%%%%%%%%%%%%%%%

\ifpreprintsty\else
\end{multicols}
\fi

\end{document}